\documentclass{icgg}

\usepackage{graphicx}     % if you want to include images
\usepackage{microtype}    % improves, among other things, line-breaks
\usepackage{balance}
 
\usepackage{amsmath}    
\usepackage{overpic,rotating,contour,subfigure}                         %f"ur \citeN,..
  
 \usepackage{graphics,subfigure}
 \usepackage{graphicx}
 \usepackage{epsfig}
   
\usepackage{amssymb}  
 
\usepackage{epsfig}  
\usepackage{psfrag,rotating}
\usepackage{amssymb,floatflt,enumerate}
\usepackage{amsmath,amscd,psfrag,leqno} 
\usepackage[mathscr]{eucal}   
\usepackage{shadethm}   
\usepackage{graphicx,subfigure}
\usepackage{overpic,contour} 
\usepackage{epstopdf}

\def\Beweisende{\square}            % Beweisende-Zeichen
\def\BewEnde{\hfill{\Beweisende}}

\def\phm{{\hphantom{-}}} % Laesst genausoviel Platz wie ein Minus!

\def\phi{\varphi}

\def\RR{{\mathbb R}}
\def\AA{{\mathbb A}}
\def\HH{{\mathbb H}}

\def\DD{{\mathbb D}}

\def\dach#1{\widehat{#1}}

\def\Vkt#1{{\mathbf #1}} 

\newcommand{\mVkt}[1]{\dach{\Vkt #1}}      % mit Dach, "Momentenvektor"
\newcommand{\dVkt}[1]{\underline{\Vkt #1}}    % unterstrichen, "dualer Vektor"

\def\Quat#1{{\frak #1}}

\def\konjQuat#1{\widetilde{\Quat#1}}
\def\kQuat#1{\konjQuat#1}

\newcommand{\go}[1]{{\sf #1}}
\newtheorem{thm}{Theorem}

\newtheorem{rem}{Remark}

\title{ALTERNATIVE INTERPRETATION OF THE  PL\"UCKER \\ QUADRIC'S AMBIENT SPACE
AND ITS APPLICATION}
 
\author{Georg NAWRATIL}
\affiliation{Vienna University of Technology, Austria}

\hypersetup{
  pdfauthor = {Georg NAWRATIL},
  pdftitle = {Alternative interpretation of the Pl\"ucker quadric's ambient space and its applications},
  pdfsubject = {ICGG 2018},
  pdfkeywords = {Instructions for formatting Full Papers, ICGG 2018, Geometry, Graphics}
}
 
\begin{document}
% Set papernumber as received upon acceptance of your contribution.
\papernumber{002}

% We need a little trickery in order to get a one-column abstract in a
% two-column text:
\twocolumn[%
\begin{@twocolumnfalse}
  \maketitle
  % Here comes the abstract...
  \begin{abstract}
    It is well-known that there exists a bijection between the set of lines of the projective 3-dimensional space $P^3$ 
		and all real points of the so-called Pl\"ucker quadric $\Psi$. Moreover one can identify each point of the 
		Pl\"ucker quadric's ambient space with a  linear complex of lines in $P^3$. 
		Within this paper we give an alternative interpretation for the points of $P^5$ as lines of an 
		Euclidean 4-space $E^4$, which are orthogonal to a fixed direction. 
		By using the quaternionic 
		notation for lines, we study straight lines in $P^5$ which correspond in the general case to cubic 2-surfaces in $E^4$.
		We show that these surfaces are geometrically connected with 
		circular Darboux 2-motions in $E^4$, as they are basic surfaces of the underlying line-symmetric motions.  
		
		Moreover we extend the obtained results to line-elements of the Euclidean 3-space $E^3$, which can be represented 
		as points of a cone over $\Psi$ sliced along the 2-dimensional generator space of ideal lines. 
		We also study straight lines of its ambient space $P^6$ and show that they correspond to ruled surface strips composed 
		of the mentioned 2-surfaces with  circles on it. 
		
		Finally we present an application of this interpretation in the context of interactive design of ruled surfaces and 
		ruled surface strips/patches based on the algorithm of De Casteljau.
  \end{abstract}
  % Don't forget to provide keywords (three to six)
\end{@twocolumnfalse}]

% Starting from here, everything is, more or less, standard LaTeX...

\section{Introduction}

Details about the following basics in line-geometry can be found in \cite{pottmann_wallner}.
Let us consider two distinct real points $\go P$ and $\go Q$ of the projective 3-space $P^3$, which possess the following homogenous 
coordinates:
\begin{equation}
(\overline{p}_0:\overline{p}_1:\overline{p}_2:\overline{p}_3),\quad 
(\overline{q}_0:\overline{q}_1:\overline{q}_2:\overline{q}_3) 
\end{equation}
with $\overline{p}_i,\overline{q}_i\in\RR$ for $i=0,\ldots ,3$. 
Then the line $\go l$ spanned by $\go P$ and $\go Q$ can be represented by the 
following  homogeneous 6-tuple 
\begin{equation}\label{pl_coord}
(l_{01}:l_{02}:l_{03}:l_{23}:l_{31}:l_{12})
\end{equation}
with 
$l_{ij}=\overline{p}_i\overline{q}_j-\overline{p}_j\overline{q}_i$.
These are the so-called Pl\"ucker coordinates of the line $\go l$. 

But contrary not each homogeneous 6-tuple corresponds to a line of $P^3$, as this set $\mathcal{L}$ of lines in $P^3$
is 4-dimensional. Only the 6-tuples fulfilling the so-called Pl\"ucker condition  
\begin{equation}\label{pk_condition}
\quad l_{01}l_{23}+l_{02}l_{31}+l_{03}l_{12}=0
\end{equation}
represent lines of $P^3$. 
Therefore there exists a bijection between the set $\mathcal{L}$ and all real points of the 
so-called Pl\"ucker\footnote{This quadric is also known as Klein quadric.}  
quadric $\Psi$ of $P^5$, which is given by Eq.\ (\ref{pk_condition}).
This bijection $\mathcal{L}\rightarrow \Psi$ is known as {\it Klein mapping}. 

\begin{rem}
Let us represent an arbitrary real point $\go U\in P^3$ by $(\overline{u}_0:\overline{u}_1:\overline{u}_2:\overline{u}_3)$. 
We identify the set of points $\go U$ determined by $\overline{u}_0=0$ with the ideal plane of 
the projective extended Euclidean 3-space. Then the 
2-dimensional generator space $L:\,\, l_{01}=l_{02}=l_{03}=0$ of $\Psi$ corresponds to the  
set of ideal lines.  \hfill $\diamond$
\end{rem}

A set $\mathcal{C}$ of lines fulfilling a linear equation 
\begin{equation}
\begin{split}
&c_{01}l_{23}+ c_{02}l_{31}+ c_{03}l_{12}+ \\
&c_{23}l_{01}+ c_{31}l_{02}+ c_{12}l_{03}= 0
\end{split}
\end{equation}
is called a linear complex of lines in $P^3$. The set $\mathcal{C}$ corresponds to the intersection of a 
hyperplane $\gamma$ and the Pl\"ucker  quadric $\Psi$. Therefore one can identify $\mathcal{C}$ with the pole $\go C$ of 
$\gamma$ with respect to $\Psi$, which has homogenous coordinates $(c_{01}:c_{02}:c_{03}:c_{23}:c_{31}:c_{12})$. 
This bijection between the points of $P^5$ and linear complexes of lines in $P^3$ is known as 
{\it extended Klein mapping}.

\subsection{Lines and line-elements of Euclidean 3-space}\label{sec:3space}

As we want to apply the theoretical results of the study at hand to the interactive design of 
rational ruled surfaces (cf.\ Sec.\ \ref{sec:app}), we restrict to the lines of the Euclidean 3-space $E^3$. 
They are represented by the real points of $\Psi\setminus L$; i.e.\ the Pl\"ucker  quadric $\Psi$ 
sliced along the 2-dimensional generator space $L$. 

The coordinates of a point $\go P\in E^3$ are given by $\Vkt p:=(p_1,p_2,p_3)$ 
with respect to the Cartesian frame $({\sf O}; x_1,x_2,x_3)$. In this case the entries of Eq.\ (\ref{pl_coord}) 
have the following geometric meaning: 
\begin{enumerate}[$\bullet$]
\item
$\Vkt l:=(l_{01},l_{02},l_{03})$ gives the direction of the line 
and differs from the zero-vector $\Vkt o$.
\item
$\mVkt l:=(l_{23},l_{31},l_{12})$ is the so-called moment-vector, which can also be computed by 
$\Vkt p\times \Vkt l$, where $\go P\in\go l$ holds. 
\end{enumerate}
Using the notation $(\Vkt l, \mVkt l)\RR$ the  Pl\"ucker condition of Eq.\ (\ref{pk_condition}) can be rewritten as 
$\langle \Vkt l,\mVkt l\rangle=0$, where $\langle\cdot,\cdot\rangle$ denotes the Euclidean scalar product. 

\begin{rem}\label{inst}
It is well known \cite{pottmann_wallner} that in $E^3$ a linear line complex $\mathcal{C}$ has the following kinematic interpretation: 
The set of lines of $\mathcal{C}$ equals the set of path-normals of an 
instantaneous motion different from the instantaneous standstill. 
For an instantaneous translation/rotation/screw motion the corresponding point $\go C\in P^5$ of the 
linear line complex $\mathcal{C}$ has the property $\go C\in L$ resp.\ $\go C\in\Psi\setminus L$ resp.\ $\go C\in P^5\setminus\Psi$.
\hfill $\diamond$
\end{rem}

For some applications (e.g.\ 3D shape recognition and reconstruction \cite{hopsw})  it is superior to 
study so-called line-elements instead of lines. As these geometric objects consist of a line $\go l$ and a 
point $\go P$ on it, we write them as $(\go l,\go P)$. Moreover we call a ruled surface together with a curve on it a 
{\it ruled surface strip}. 

According to \cite{opw} the Pl\"ucker coordinates of lines 
can be extended for line-elements of $E^3$ by:
\begin{equation}\label{hierl}
(l_{01}:l_{02}:l_{03}:l_{23}:l_{31}:l_{12}:l)
\end{equation}
with
$l:=\langle \Vkt p,\Vkt l\rangle$. In the remainder of the article we abbreviate this 
homogenous 7-tuple by $(\Vkt l, \mVkt l, l)\RR$. 
Obviously there is a bijection between the set of line-elements of $E^3$ 
and all real points of $P^6$ located on  a cone $\Lambda$ over $\Psi$, which is sliced along the 3-dimensional 
generator space $G:\,\, l_{01}=l_{02}=l_{03}=0$ of $\Lambda$.
Again the points $(\Vkt c, \mVkt c, c)\RR$ of $\Lambda$'s ambient space $P^6$ can be interpreted as linear complexes of 
line-elements; i.e. the set of line-elements $(\Vkt l, \mVkt l, l)\RR$ fulfilling the linear equation
\begin{equation}
\langle \Vkt c,\mVkt l\rangle + \langle \mVkt c,\Vkt l\rangle + cl =0.
\end{equation}

\begin{rem}
For reasons of completeness it should be noted 
that the set of line-elements of $P^3$ is studied in \cite{boris} 
and \cite[Sec.\ 6]{nawratil_line_element}, respectively. 
Moreover the set of oriented line-elements is investigated in \cite{nawratil_survey}. 
\hfill $\diamond$
\end{rem}

\subsection{Outline}

In Sec.\ \ref{sec:line4} we give an alternative interpretation for the points 
of $\Psi$'s ambient space $P^5$ as the set of lines of the Euclidean 
4-space $E^4$, which are orthogonal to a fixed direction. 
The obtained results are extended to line-elements in Sec.\ \ref{sec:lineele4}. 
In Sec.\ \ref{sec:straight} we study 2-surfaces (resp.\ surface strips) of $E^4$, which correspond to 
straight lines in $P^5$ (resp.\ $P^6$).  
The connection of these 2-surfaces with circular Darboux 2-motions of $E^4$ is discussed in  Sec.\ \ref{sec:linesym}. 
We close the paper (cf.\ Sec.\ \ref{sec:app}) with an application of the given interpretation in the context of 
interactive design of ruled surfaces and ruled surface strips/patches based on the 
algorithm of De Casteljau.

\section{Lines in Euclidean 4-space}\label{sec:line4}

The quaternionic representation of lines in $E^4$ allows a very compact notation. This 
formulation, which is shortly repeated in the next subsection, was already used by the author 
in \cite{nawratil_clifford} and \cite[Sec.\ 6]{nawratil_line_element}, respectively. 

\subsection{Quaternionic representation}

$\frak{Q}:=q_0+q_1\Vkt i+q_2\Vkt j+q_3\Vkt k$ with $q_0,\ldots,q_3\in\RR$ is an element of the skew field of quaternions $\HH$, 
where $\Vkt i,\Vkt j,\Vkt k$ are the quaternion units. The scalar part is $q_0$ and the pure part equals $q_1\Vkt i+q_2\Vkt j+q_3\Vkt k$, 
which is also denoted by $\frak{q}$. 
The conjugated quaternion to  $\frak{Q}=q_0+\frak{q}$ is given by 
$\widetilde{\frak{Q}}:=q_0-\frak{q}$ and $\frak{Q}$ is called a unit-quaternion for
$\frak{Q}\circ \widetilde{\frak{Q}}=1$,
where $\circ$ denotes the 
quaternion multiplication.  

We embed points $\go P$ of $E^4$ with coordinates $(p_0,p_1,p_2,p_3)$ 
with respect to the Cartesian frame $({\sf{O}} ; x_0,x_1,x_2,x_3)$ into the set of quaternions by 
the mapping $\iota:\RR^4\rightarrow \HH$ with 
\begin{equation}\label{einbettung4}
\begin{split}
(p_0,p_1,p_2,p_3)\,\mapsto\, \frak{P}:&=p_0+p_1\Vkt i+p_2\Vkt j+p_3\Vkt k \\
&=p_0+\Quat{p}.
\end{split}
\end{equation}
Let us identify $E^3$ with the hyperplane $x_0=0$.

The so-called {\it homogenous minimal coordinates} (cf.\ \cite[Def.\ 3]{nawratil_line_element})
of a line $\go l\in E^4$ can be written in terms of quaternions as
\begin{equation}\label{Lundm}
(\Quat{L},\Quat{m})\RR 
\end{equation}
with $\Quat{m}:=\kQuat{L}\circ\Quat{F}$, where the latter quaternion results from the embedding $\iota$ of the 
pedal point $\go F$ of the line $\go l$ with respect to the origin $\go O$ of the reference frame. 
Note that $\Quat m$ is a pure quaternion and that $\Quat{L}$ correspond to the direction of the line $\go l\in E^4$. 

There is a bijection between the set of lines of $E^4$ and the points of $P^6$, which is sliced along the 
2-space $l_0=l_1=l_2=l_3=0$; i.e.\ $\Quat{L}=0$. 

\subsection{Alternative Interpretation}\label{sec:altint}

In the following we are only interested in the subset $\mathcal{M}$ of lines of $E^4$, which are orthogonal to 
the $x_0$-direction. As a consequence $\Quat{L}$ has to be a pure quaternion, i.e.\ the  {\it homogenous minimal coordinates} 
of a line $\go l\in\mathcal{M}$ read as:
\begin{equation}
(\Quat{l},\Quat{m})\RR. 
\end{equation}
Thus there is a bijection between the set  $\mathcal{M}$ and the points of $P^5$, which is sliced along the 
2-space $L:\,\,l_1=l_2=l_3=0$; i.e.\ $\Quat{l}=0$. Moreover lines of $\mathcal{M}$ belonging to $E^3$ 
(given by $x_0=0$) are determined by the fact that $\Quat{F}$, which 
can be computed as  
\begin{equation}\label{berC}
\tfrac{\Quat{l}\circ\Quat{m}}{\Quat{l}\circ\kQuat{l}}, 
\end{equation}
is a pure quaternion. 
Therefore 
\begin{equation}
f_0=\tfrac{1}{2}\tfrac{\Quat{l}\circ\Quat{m}+\kQuat{m}\circ\kQuat{l}}{\Quat{l}\circ\kQuat{l}} \\
=\tfrac{1}{2}\tfrac{\Quat{l}\circ\Quat{m}+\Quat{m}\circ\Quat{l}}{\Quat{l}\circ\kQuat{l}}
\end{equation}
holds and the condition $f_0=0$ is equivalent with
\begin{equation}
l_1m_1+l_2m_2+l_3m_3=0, 
\end{equation}
which is exactly the Pl\"ucker condition given in Eq.\ (\ref{pl_coord}). 
This completes the alternative interpretation of $P^5\setminus L$, i.e.\ the 
ambient space of the Pl\"ucker quadric $\Psi$ sliced along the 2-dimensional generator space $L$.

\subsection{Projection on the Pl\"ucker quadric} 

According to the extended Klein mapping and Rem.\ \ref{inst} 
every point $(\Vkt c,\mVkt c)\RR$ of $P^5$, which is not located on $\Psi$, 
corresponds to the path-normals of a instantaneous screw motion. 
The Pl\"ucker coordinates $(\Vkt a,\mVkt a)\RR$ of the 
so-called axis of this instantaneous screw motion 
can be computed according to \cite[Thm.\ 3.1.9]{pottmann_wallner} as
\begin{equation}\label{axis}
\Vkt a=\Vkt c, \qquad
\mVkt a=\mVkt c-\tfrac{\langle \Vkt c, \mVkt c\rangle}{\langle \Vkt c, \Vkt c\rangle}\Vkt c.
\end{equation} 
For points $(\Vkt c,\mVkt c)\RR$ of $\Psi\setminus L$ Eq.\ (\ref{axis}) simplifies to 
\begin{equation}
\Vkt a=\Vkt c, \qquad \mVkt a=\mVkt c,
\end{equation}
which is the axis of the instantaneous rotation. 
In total Eq.\ (\ref{axis}) implies a mapping $\mu$: $P^5\setminus L\rightarrow \Psi\setminus L$ with 
$(\Vkt c,\mVkt c)\RR \mapsto (\Vkt a,\mVkt a)\RR$.

In analogy\footnote{In this context see also \cite{swc} and \cite[Rem.\ 10]{nawratil_survey}.}
to \cite[Thm.\ 1]{without_study} the fibers of this mapping $\mu$ can be written as 
follows: 
\begin{thm}\label{thm1}
The fiber of point $(\Vkt c,\mVkt c)\RR \in P^5\setminus L$ with respect to the mapping $\mu$ 
is a straight line through $(\Vkt c,\mVkt c)\RR$ that intersects the 2-dimensional generator space $L$ 
in the point $(\Vkt o,\Vkt c)\RR$. 
\end{thm}

What is the geometric meaning of the mapping $\mu$ in terms of our alternative interpretation?
In order to clarify this we rewrite the mapping $\mu$ in the quaternionic formulation; i.e.\  
\begin{equation}\label{mu_quat}
\mu:\,\, (\Quat{l},\Quat{m})\RR \mapsto
(\Quat{l},\Quat{m}+\tfrac{1}{2}\tfrac{\Quat{l}\circ\Quat{m}+\Quat{m}\circ\Quat{l}}{\Quat{l}\circ\kQuat{l}}\Quat{l})\RR
\end{equation}
and compute the pedal point $\go F^*$ of $\go l^*$ given by $(\Quat{l}^*,\Quat{m}^*)\RR:=\mu(\Quat{l},\Quat{m})\RR$.
The formula of Eq.\ (\ref{berC}) implies 
\begin{equation}
\begin{split}
\Quat{F}^*&=\tfrac{\Quat{l}^*\circ\Quat{m}^*}{\Quat{l}^*\circ\kQuat{l}^*} \\
&=\tfrac{\Quat{l}\circ\Quat{m}}{\Quat{l}\circ\kQuat{l}} - 
\tfrac{1}{2}\tfrac{\Quat{l}\circ\Quat{m}+\Quat{m}\circ\Quat{l}}{\Quat{l}\circ\kQuat{l}} \\
&=\Quat{F}-f_0.
\end{split}
\end{equation} 
This shows that the mapping $\mu$ corresponds to the orthogonal projection of the line $\go l\in\mathcal{M}$ onto 
$E^3$ (given by $x_0=0$). We denote this orthogonal projection $E^4\rightarrow E^3$ by $\pi$.

\section{Lines-elements in Euclidean 4-space}\label{sec:lineele4}
In this section we extend the results of Sec.\ \ref{sec:line4} to the 
line-elements of $E^4$.

The so-called {\it homogenous minimal coordinates} (cf.\ \cite[Def.\ 4]{nawratil_line_element})
of a line-element can be written in terms of quaternions as
\begin{equation}
(\Quat{L},l+\Quat{m})\RR 
\end{equation}
with $\Quat{L}$ and $\Quat{m}$ of Eq.\ (\ref{Lundm}) and $l$ of Eq.\ (\ref{hierl}). 
There is a bijection between the set of line-elements of $E^4$ and the points of $P^7$, which is sliced along the 
3-space $l_0=l_1=l_2=l_3=0$; i.e.\ $\Quat{L}=0$.

The subset of line-elements 
$(\go l,\go P)$ of $E^4$, where $\go l$ is orthogonal to the $x_0$-direction, is denoted by $\mathcal{N}$. 
As a consequence $\Quat{L}$ has to be a pure quaternion, i.e.\ the  {\it homogenous minimal coordinates} 
of a line-element $(\go l,\go P)\in\mathcal{N}$ read as:
\begin{equation}
(\Quat{l},l+\Quat{m})\RR. 
\end{equation}
Analog considerations to Sec.\ \ref{sec:altint} imply the following statement: 
There is a bijection between the set  $\mathcal{N}$ and the points of $P^6$, which is sliced along the 
3-space $G:\,\, l_1=l_2=l_3=0$; i.e.\ $\Quat{l}=0$. Moreover line-elements of $\mathcal{N}$ belonging to $E^3$ 
(given by $x_0=0$) are located on the cone $\Lambda\setminus G$.

We can also extend the projection $\mu$ of Eq.\ (\ref{mu_quat}) to the set $\mathcal{N}$ by
\begin{equation}  
\begin{split}
\nu:\,\, &P^6\setminus G \rightarrow \Lambda\setminus G \\ 
&(\Vkt c,\mVkt c,c)\RR \mapsto (\Vkt a,\mVkt a,c)\RR.
\end{split}
\end{equation}
In analogy to Thm.\ \ref{thm1} we get:
\begin{thm}\label{thm2}
The fiber of point $(\Vkt c,\mVkt c,c)\RR \in P^6\setminus G$ with respect to the mapping $\nu$ 
is a straight line through $(\Vkt c,\mVkt c,c)\RR$ that intersects the 3-dimensional generator space $G$ 
in the point $(\Vkt o,\Vkt c,0)\RR$. 
\end{thm}
The quaternionic notation of the mapping $\nu$ is given by
\begin{equation}
\nu:\,\, (\Quat{l},l+\Quat{m})\RR \mapsto
(\Quat{l},l+\Quat{m}+\tfrac{1}{2}\tfrac{\Quat{l}\circ\Quat{m}+\Quat{m}\circ\Quat{l}}{\Quat{l}\circ\kQuat{l}}\Quat{l})\RR
\end{equation}
and the geometric meaning of $\nu$ in terms of our alternative interpretation reads as follows:
The mapping $\nu$ corresponds to the orthogonal projection $\pi$ of the line-element  $(\go l,\go P)\in\mathcal{N}$ onto 
$E^3$  (given by $x_0=0$).

\subsection{Lower-dimensional analogue}
\begin{figure}[t]
\begin{center} 
\begin{overpic}
    [height=23mm]{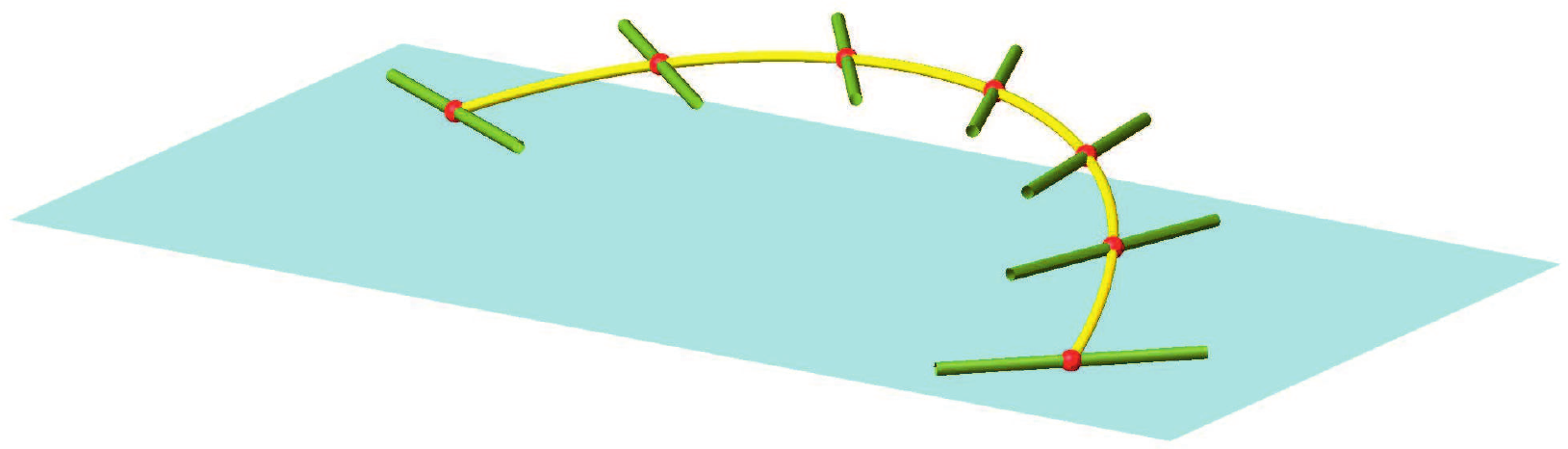}
		\begin{footnotesize}
		\put(6.5,14.5){$x_1x_2$-plane}
\end{footnotesize}         
  \end{overpic} 
\end{center} 
\caption{The ruled surface strip is conoidal as the generators are parallel to the $x_1x_2$-plane.} 
\label{fig1}
\end{figure}

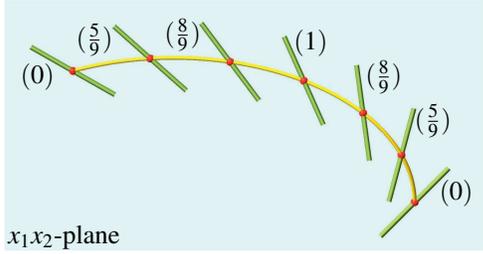
\begin{figure}[t]
\begin{center} 
 \begin{overpic}
    [height=35mm]{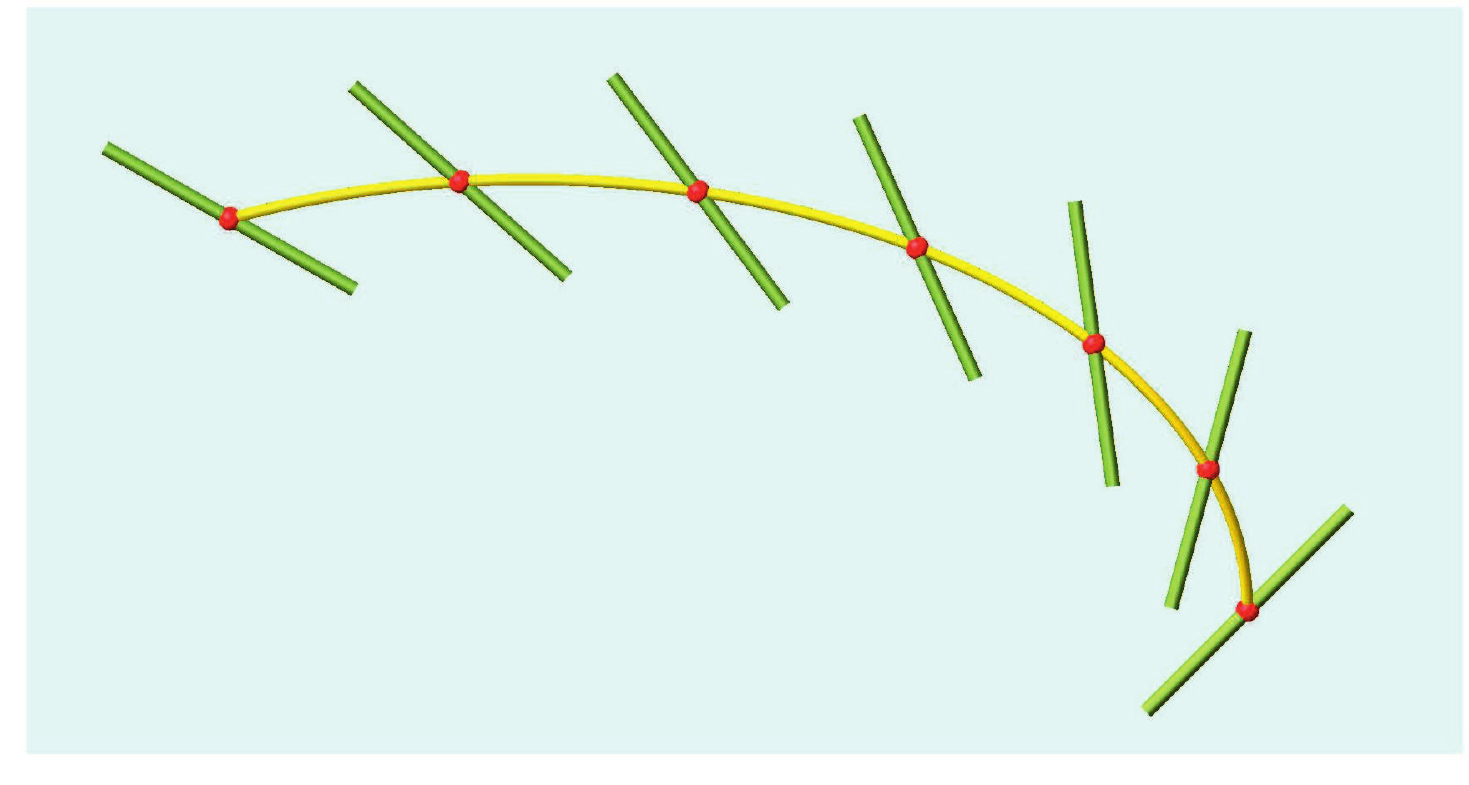}
\begin{footnotesize}
\put(90,12.5){$(0)$}
\put(85,27){$(\tfrac{5}{9})$}
\put(75,36){$(\tfrac{8}{9})$}
\put(60.5,43){$(1)$}
\put(34.2,44.5){$(\tfrac{8}{9})$}
\put(16.3,43.5){$(\tfrac{5}{9})$}
\put(5,36){$(0)$}
\put(2.5,3.5){$x_1x_2$-plane}
\end{footnotesize}      
  \end{overpic} 
\end{center} 
\caption{The image of the conoidal surface strip of Fig.\  \ref{fig1} under $\eta$ is obtained by considering the 
top view. In addition we label the line-elements in the top view by the 
$x_3$-coordinate. In German such a map is known as "{\it kotierte Projektion}".
 } 
\label{fig2}
\end{figure}

In order to make the things more descriptive we stress the following lower-dimensional analogue: 
Consider the set $\mathcal{Q}$ of line-elements of $E^3$, those lines are orthogonal to the $x_3$-direction. 
Moreover we consider the set $\mathcal{P}$ of line-elements, which are contained in the $x_1x_2$-plane. 
This implies the relation  $\mathcal{P}\subset\mathcal{Q}$.
If we apply an orthogonal projection $\eta$ along the $x_3$-direction on the $x_1x_2$-plane (analogue of $\pi$)
to line-elements of  $\mathcal{Q}$, we obtain line-elements of $\mathcal{P}$. This is illustrated in Figs.\ 
\ref{fig1} and \ref{fig2}.

\begin{rem}
By omitting the point of the line-element we get the lower-dimensional analogue 
for the situation discussed in Sec.\ \ref{sec:line4}. \hfill $\diamond$
\end{rem}

\section{Straight lines}\label{sec:straight}

In the following we study the geometric objects, which are defined by straight lines 
in  $P^5\setminus L$ and  $P^6\setminus G$, respectively.

\subsection{Straight lines in $P^5\setminus L$}\label{sec:straight1}

Given are two distinct lines $\go l_1,\go l_2\in\mathcal{M}$ with quaternionic representation  
\begin{equation}
(\Quat l_i,\Quat m_i)\RR \quad\text{with}\quad i=1,2.
\end{equation}
Moreover we consider the line $\go g$ spanned by the corresponding points in $P^5$, which is given by
\begin{equation}\label{geradeg}
[t(\Quat l_1,\Quat m_1)+(1-t)(\Quat l_2,\Quat m_2)]\RR \quad \text{with}\quad t\in\RR.
\end{equation}
 Now we can distinguish the following cases: 
\begin{enumerate}[1.]
\item
$\go l_1$ and $\go l_2$ are located within a plane. Then it can easily be seen that 
$\go g$ corresponds to the pencil of lines spanned by $\go l_1$ and $\go l_2$. 
\item 
$\go l_1$ and $\go l_2$ are skew; i.e. they span a 3-space. 
The surface $\Gamma\in E^4$ which corresponds to $\go g$ is studied next.
\end{enumerate}

\subsubsection{Properties of $\Gamma$}

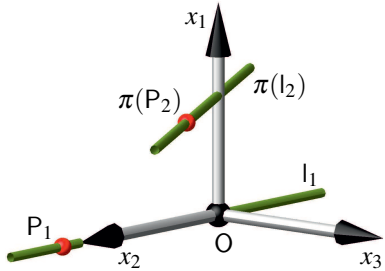
\begin{figure}[b]
\begin{center} 
 \begin{overpic}
    [height=35mm]{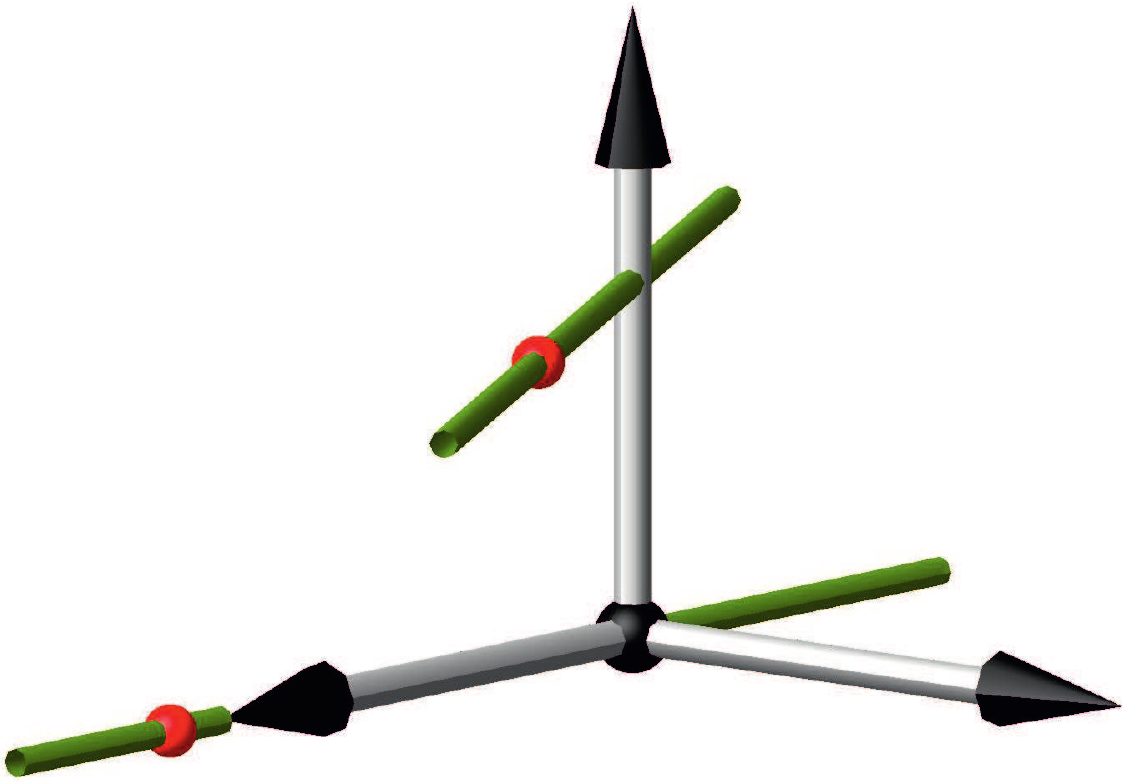}
\begin{footnotesize}
\put(65,45){$\pi(\go l_2)$}
\put(29.5,41){$\pi(\go P_2)$}
\put(5.5,7.5){$\go P_1$}
\put(55,3){$\go O$}
\put(78,21.5){$\go l_1$}
\put(47.5,63){$x_1$}
\put(30,0){$x_2$}
\put(93.5,0){$x_3$}
\end{footnotesize}      
  \end{overpic} 
\end{center} 
\caption{$\pi(\go l_2)$ intersects the $x_1$-axis at height $n$ and encloses with the $x_2$-axis the angle $\alpha\in]0^{\circ},180^{\circ}[$. 
 } 
\label{fig3}
\end{figure}

Clearly, $\Gamma$ is a 2-surface in $E^4$ as it is generated by a 1-parametric set of lines. 
As all these lines belong to $\mathcal{M}$, we can say that $\Gamma$ is a conoidal 2-surface with 
respect to the director hyperplane $x_0=0$. For the study of further properties of $\Gamma$ we 
use the following coordinatisation: 

Without loss of generality we can assume that 
\begin{enumerate}[$\bullet$]
\item
$\go l_1$ is located in $E^3$ (given by $x_0=0$), 
\item
$\go l_2$ belongs to the hyperplane $x_0=h$ with $h\in\RR$, 
\item
$\go l_1$ coincides with the $x_2$-axis, 
\item
$\pi(\go l_2)$ is parallel to the $x_2x_3$-plane, 
\item
$x_1$-axis is the common normal of $\go l_1$ and $\pi(\go l_2)$. 
\end{enumerate}
This choice of the coordinate system, which is illustrated in Fig.\ \ref{fig3}, yields:
\begin{align}
\Quat l_1&=\Vkt j, &\quad \Quat l_2&=\cos{\alpha}\Vkt j+\sin{\alpha}\Vkt k, \\
\Quat F_1&=0,  &\quad \Quat F_2&=h+n\Vkt i, 
\end{align}
implying $\Quat m_1=0$ and 
\begin{equation*}
\Quat m_2= (h\cos{\alpha}-n\sin{\alpha}) + (h\sin{\alpha}+n\cos{\alpha})\Vkt i,
\end{equation*}
which can be inserted into Eq.\ (\ref{geradeg}). 

According to Eq.\ (\ref{berC}) the pedal point of $\go l\in\Gamma$ (in dependency of $t$) reads as
\begin{equation}\label{eqf}
\Vkt s:=
\begin{pmatrix}
\frac{(t-1)(th\cos{\alpha}-tn\sin{\alpha}-th+h)}{2(t\cos{\alpha}-t)(t-1)-1} \\
\frac{(t-1)(th\sin{\alpha}+tn\cos{\alpha}-tn+n)}{2(t\cos{\alpha}-t)(t-1)-1} \\
0 \\
0
\end{pmatrix}
\end{equation} 
and the direction $\Vkt r$ of the ruling $\go l\in\Gamma$ (in dependency of $t$) equals
\begin{equation}
\Vkt r:=
\begin{pmatrix}
0 \\ 0 \\
\frac{t+(1-t)\cos{\alpha}}{2(t\cos{\alpha}-t)(t-1)-1} \\
\frac{(1-t)\sin{\alpha}}{2(t\cos{\alpha}-t)(t-1)-1}. 
\end{pmatrix}.
\end{equation}
Therefore $\Gamma\in E^4$ can be rationally parametrized as follows 
with respect to the parameters $t$ and $u$:
\begin{equation}\label{eqgam}
\Gamma:\,\, \Vkt g=\Vkt s + u\Vkt r \quad \text{with} \quad u\in\RR.
\end{equation}

\begin{thm}
$\Gamma$ possesses a rational quadratic parametrization and it is a so-called LN-surface.
\end{thm}

\noindent
{\sc Proof:}
LN-surfaces in $E^4$ are discussed in \cite{martin} and can be characterized as follows: 
For any three-space of $E^4$ there exists a unique tangent plane of $\Gamma$ parallel to it. 
According to \cite[Eq.\ (12)]{martin} this is equivalent with the existence of 
a rational solution of the two equations
\begin{equation}
\langle\Vkt w,\tfrac{\partial}{\partial t}\Vkt g\rangle =0 \quad \text{and}\quad
\langle\Vkt w,\tfrac{\partial}{\partial s}\Vkt g\rangle =0 
\end{equation}
for $t$ and $u$ in dependence of  
$\Vkt w\in \RR^4$. 
It can easily be checked that this criterion is fulfilled. \hfill $\BewEnde$

\begin{thm}
$\Gamma$ is a cubic conoidal 2-surface. 
\end{thm}

\noindent
{\sc Proof:}
We consider Eq.\ (\ref{eqgam}) coordinate-wise 
and eliminate from these four equations the 
parameters $t$ and $u$ by resultant method. 
This yields the following set of equations:
\begin{equation}\label{set}
\begin{split}
(a)\phm&\sin{\alpha}(g_1g_2^2+g_1g_3^2-hg_3g_2-ng_3^2) \\&-g_3\cos{\alpha}(ng_2-hg_3)=0, \\
(b)\phm&\sin{\alpha}(hg_3^2-ng_3g_2-g_0g_3^2-g_0g_2^2) \\&+g_3\cos{\alpha}(hg_2+ng_3)=0,\\
(c)\phm&\sin{\alpha}(g_0^2-ng_1+g_1^2-hg_0) \\&-\cos{\alpha}(ng_0-g_1h)=0. 
\end{split}
\end{equation}
In order to obtain the projective closure $P^4$ of $E^4$, we 
homogenize the coordinates by $g_i=\tfrac{v_i}{v}$, for $i=0,\ldots ,3$  
whereby $v$ denotes the homogenizing variable. 
The degree of $\Gamma\in P^4$ corresponds to the number of intersection points (counted by multiplicity) of 
this 2-surface with a plane. We choose this plane as follows:
\begin{equation}
v_0=v_2+v_3+v, \quad v_1=v_2-v_3-v.
\end{equation} 
Under consideration of these relations
the three equations (a,b,c) of  Eq.\ (\ref{set}) 
are homogenous equations in $v,v_2,v_3$. We eliminate from equation (i) and (j) the 
unknown $v$ by resultant method for pairwise distinct $i,j\in\left\{a,b,c\right\}$, 
which yields $Res(i,j)$. The greatest common divisor of $Res(a,b)$, $Res(a,c)$ and 
$Res(b,c)$ equals $\sin{\alpha}(v_2^2+v_3^2)V$ with
\begin{equation*}
V:=\sin{\alpha}(2v_2+hv_3+nv_3)+v_3\cos{\alpha}(n-h).
\end{equation*} 
This yields one real and two conjugate complex intersection points (with multiplicty one), thus 
we can conclude that $\Gamma$ is \bigskip cubic. \hfill $\BewEnde$

Finally, it should be noted (cf.\  \cite[Sec.\ 2]{ge_ravani}) that the image of $\Gamma$ under $\pi$ is the 
{\it Pl\"ucker conoid}, which is also  known as {\it cylindroid}.

\subsection{Straight lines in $P^6\setminus G$}

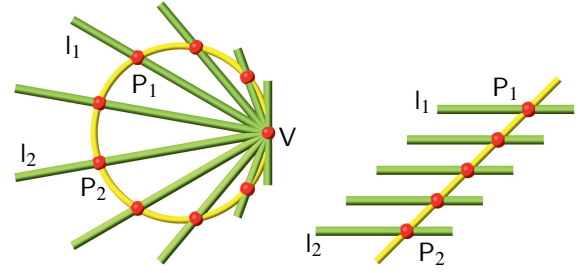
\begin{figure}[t]
\begin{center} 
 \begin{overpic}
    [height=35mm]{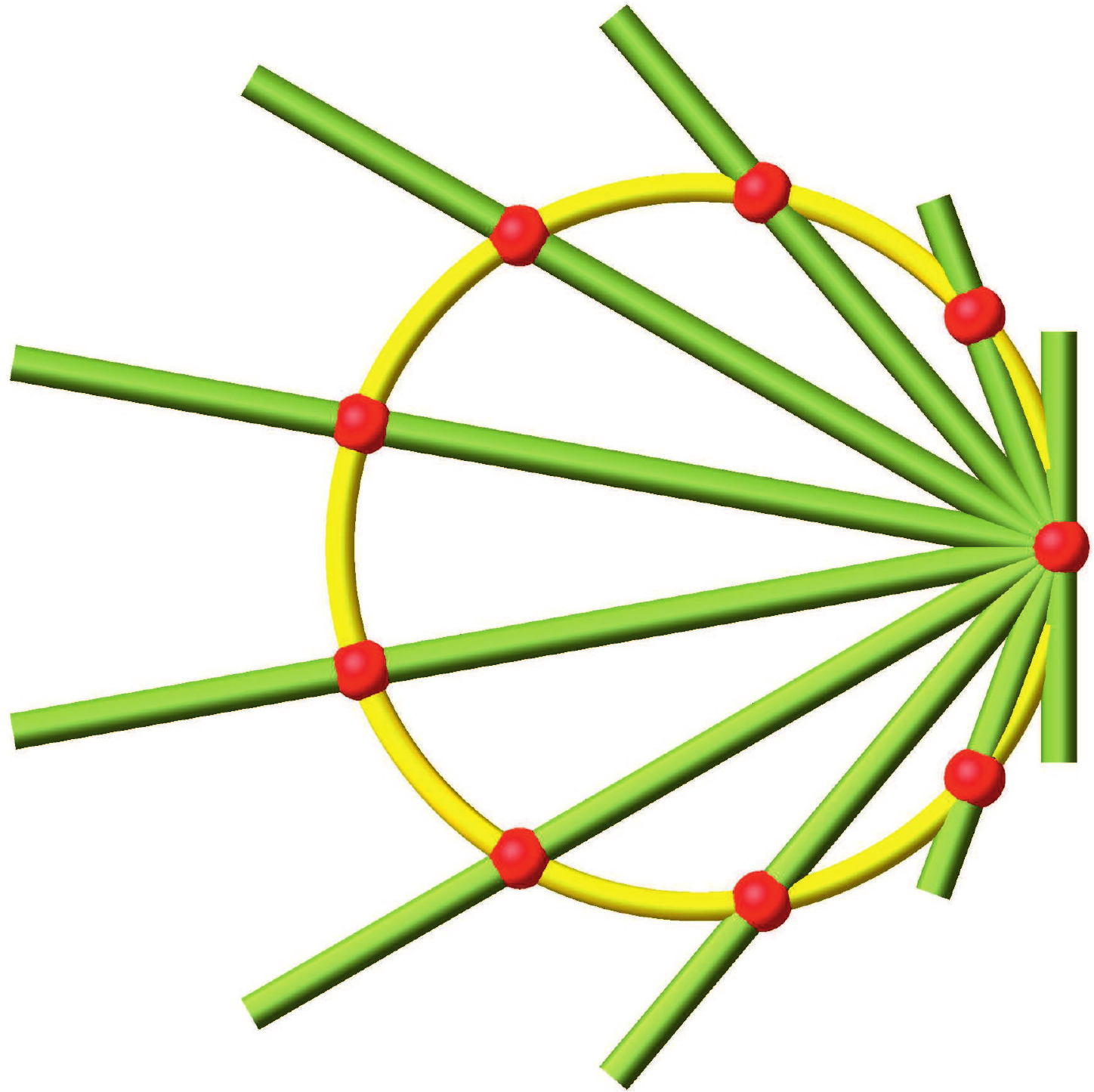}
\begin{footnotesize}
\put(101,45){$\go V$}
\put(45,65){$\go P_1$}
\put(25.5,26){$\go P_2$}
\put(20,80){$\go l_1$}
\put(3,39){$\go l_2$}
\end{footnotesize}      
  \end{overpic} 
	\quad
	\begin{overpic}
    [height=25mm]{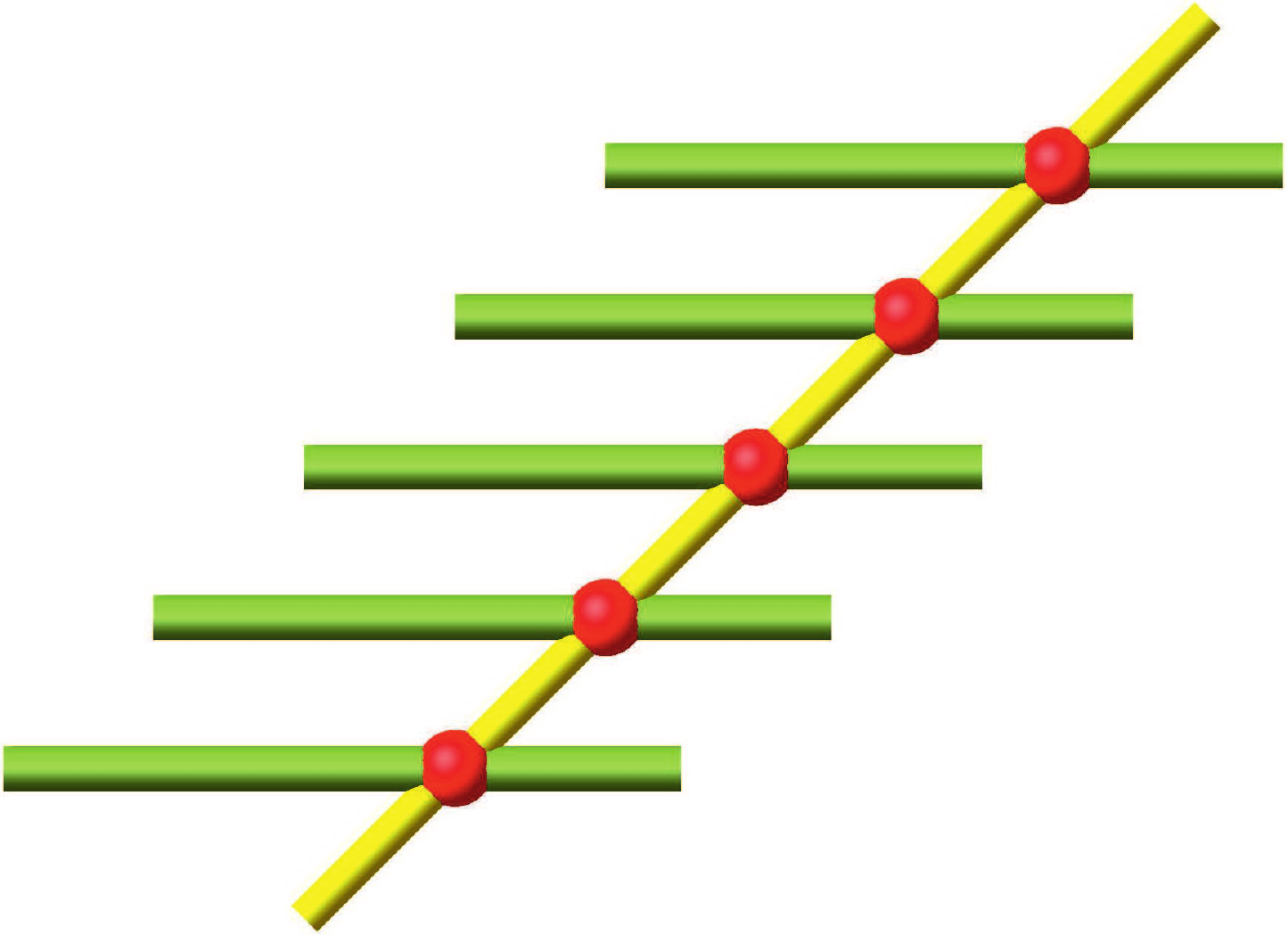}
\begin{footnotesize}
\put(68,65){$\go P_1$}
\put(39.5,1.5){$\go P_2$}
\put(38,60){$\go l_1$}
\put(-4,5){$\go l_2$}
\end{footnotesize}      
  \end{overpic} 
\end{center} 
\caption{Right/Left: The special ruled surface strips of item (1a)/(1b). 
 } 
\label{fig4}
\end{figure}

Now we consider  two distinct line-elements $(\go l_i,\go P_i)\in\mathcal{N}$ with 
quaternionic representation
\begin{equation}
(\Quat l_i,l_i+\Quat m_i)\RR \quad\text{with}\quad i=1,2
\end{equation}
and the straight line $\go q$ in  $P^6\setminus G$ spanned by the corresponding points in $P^6$; i.e.\ 
\begin{equation}\label{geradegplus}
[t(\Quat l_1,l_1+\Quat m_1)+(1-t)(\Quat l_2,l_2+\Quat m_2)]\RR.
\end{equation}
If the underlying lines $\go l_1$ and $\go l_2$ are
\begin{enumerate}[1.]
\item
{\bf coplanar}, then we can distinguish two cases:
	\begin{enumerate}[(a)]
	\item
	$\go l_1\parallel\go l_2$: In this case $\go q$ corresponds to a ruled surface strip which consists of a 
	parallel line pencil (spanned by $\go l_1$ and $\go l_2$) with a line on it  (cf.\ Fig.\ \ref{fig4}, right). 
	
	For the special case $\go l_1=\go l_2$ 
	the line $\go q$ corresponds to the set of line-elements which have the same carrier line $\go l_1=\go l_2$. 
	\item
	$\go l_1\nparallel\go l_2$: In this case $\go q$ corresponds to a ruled surface strip consisting of a 
	line pencil, where the vertex $\go V$ is the intersection point of $\go l_1$ and $\go l_2$, and a circle on it, 
	which is determined by $\go V,\go P_1,\go P_2$ (cf.\ Fig.\ \ref{fig4}, left).
	\end{enumerate} 
The statements given in item (1a) are trivial (proofs are left to the reader) and those of 
item (1b) follow from \cite[Cor.\ 2]{opw}. 
\item 
{\bf skew}, then $\go g$ corresponds to a ruled surface strip $(\Gamma,\go k)$ with the underlying ruled surface $\Gamma$ of Sec.\ \ref{sec:straight1}. 
Therefore we are only left with the question for the curve $\go k$ on $\Gamma$, which is answered next.
\end{enumerate}

 \begin{thm}\label{thmcircles}
$\go k$ is a circle, which implies that $\Gamma$ carries a 2-parametric set of circles.
\end{thm}

\noindent
{\sc Proof:}
The curve $\go k$ on $\Gamma$ is given by 
\begin{equation}\label{parkreis}
\Vkt k=\Vkt s + (tl_1+(1-t)l_2)\Vkt r.
\end{equation}
As this is a rational quadratic parametrization in $t$, the curve $\go k$ has to be 
a conic section in $E^4$. 
By introducing homogenous coordinates  
$k_i=\tfrac{q_i}{q}$ for $i=0,\ldots ,3$, 
we are able to intersect the curve $\go k$ with the 
hyperplane at infinity determined by $q=0$. 
The corresponding equation 
\begin{equation}
2(t\cos{\alpha}-t)(t-1)-1=0
\end{equation}
has the solutions 
$t=\tfrac{1-\cos{\alpha}\pm I \sin{\alpha}}{2-2\cos{\alpha}}$
where $I$  denotes the complex unit. Direct computation shows that the corresponding two 
intersection points are located on the absolute sphere $q_0^2+q_1^2+q_2^2+q_3^2=0$ ($\Rightarrow$ $\go k$ is a circle). 

$\Gamma$ carriers a 2-parametric set of circles as $l_1$ and $l_2$ can take arbitrary real values. 
\hfill $\BewEnde$

\begin{thm}
The striction curve $\go s$ of $\Gamma$ is a circle. 
Moreover  $\go s$ is a geodesic curve of $\Gamma$.
\end{thm}

\noindent
{\sc Proof:}
According to \cite[p.\ 364]{pottmann_wallner} the {\it striction point} of a ruling is the point of contact of the 
tangent plane, which is orthogonal to the asymptotic plane ($=$ tangent plane in the ruling's 
ideal point). Moreover it should be noted (cf.\ \cite[p.\ 11]{plass}) 
that the tangent planes along a ruling of a 2-surface in $E^4$ are located within a hyperplane. 

The asymptotic plane is spanned by the vectors $\Vkt r$ and $\tfrac{\partial}{\partial t}\Vkt r$. It can easily be checked that the 
tangent plane along $\Vkt s$ of Eq.\ (\ref{eqf}), which is spanned by $\Vkt r$ and  $\tfrac{\partial}{\partial t}\Vkt s$, is 
orthogonal to the asymptotic plane. Therefore the striction curve  $\go s$ is given by Eq.\ (\ref{eqf}). 
Moreover $\go s$ is a circle, which follows from Thm.\ \ref{thmcircles} by setting $l_1=l_2=0$ in Eq.\ (\ref{parkreis}).

According to \cite[p.\ 14]{plass} the following statement holds: If a curve $\go s$ on a ruled 2-surface in $E^4$ possesses two 
of the following properties, it possesses the third:  
\begin{enumerate}
\item
$\go s$ is the striction curve,
\item
$\go s$ is a geodesic,
\item
$\go s$ cuts the rulings with a fixed angle.
\end{enumerate}
It can easily be verified that $\go s$ intersects all rulings orthogonal. Therefore $\go s$ is a \bigskip geodesic. 
\hfill $\BewEnde$

Finally it should be noted that $\pi(\go s)$ coincides with the common normal of $\go l_1$ and $\pi(\go l_2)$. 
All other circles $\go k$ on $\Gamma$ are mapped to ellipses $\pi(\go k)$.

\subsection{Kinematic relevance of $\Gamma$}\label{sec:linesym}

Now we want to study the one-parametric motion in $E^4$, which is generated by reflecting the coordinate frame 
in the one-parametric set of $\Gamma$'s rulings. Such a motion is called line-symmetric and $\Gamma$ is 
the corresponding {\it basic surface} (cf.\ \cite[Chap.\ 9,\S 7]{bottemaroth}). 

Without loss of generality (cf.\ Sec.\ \ref{sec:straight1}) 
we can assume that the rulings $\go l\in\Gamma$ with quaternionic representation $(\Quat l,\Quat m)\RR$ 
have the direction $(0,0,\cos{\beta},\sin{\beta})^T$ for $\beta\in[0^{\circ},180^{\circ}[$. 
Moreover we consider the orthogonal vectors 
\begin{equation}\label{eqhs}
\Vkt h_1=\begin{pmatrix} 1 \\0 \\0 \\ 0\end{pmatrix}, \,\,
\Vkt h_2=\begin{pmatrix} 0 \\1 \\0 \\ 0\end{pmatrix}, \,\,
\Vkt h_3=\begin{pmatrix} 0 \\0 \\-\sin{\beta} \\ \cos{\beta}\end{pmatrix}. 
\end{equation}
Then the reflection $\kappa$ in $\go l$ can be replaced by the composition of three reflections $\rho_i$ in 
the hyperplanes $H_i$ through $\go l$, which are orthogonal to $\Vkt h_i$ of Eq.\ (\ref{eqhs}) 
for $i=1,2,3$. Therefore it is clear that $\kappa$ is a orientation-reversing isometry. 

But we want to describe the obtained one-parametric motion in $E^4$ by 
orientation-preserving isometries ($=$ displacements). Therefore we first reflect the coordinate frame in the hyperplane $x_0=0$ 
(reflexion $\rho_0$) and then we apply $\kappa$. In total a point $\go P$ of $E^4$ is transformed by
\begin{equation} 
\kappa(\rho_0(\go P))=\rho_3(\rho_2(\rho_1(\rho_0(\go P)))).
\end{equation}
As $\rho_1(\rho_0(\go P))$ is a composition of two reflexions in parallel hyperplanes it equals a pure 
translation orthogonal to these hyperplanes. We denote this translation along the  $x_0$-axis by $\tau$.
Therefore we only have to clarify the action of the composition of $\rho_3$ and $\rho_2$ on $\tau(\go P)$.
As the hyperplanes $H_2$ and $H_3$ intersect orthogonally along a plane $T$ through $\go l$, which is parallel to the $x_0$-axis, 
the composition of $\rho_3$ and $\rho_2$ equals the rotation $\delta$ about $T$ by $180^{\circ}$ ($=$ reflexion in $T$). 
This yields
\begin{equation} 
\kappa(\rho_0(\go P))=\delta(\tau(\go P)).
\end{equation}
It can easily be verified by direct computations that this displacement can be written in 
quaternionic notation as follows: 
\begin{equation}\label{schoenflies4}
\frak P\,\,\mapsto\,\, \frak l\circ\frak P \circ \widetilde{\frak{l}} - 2\frak l\circ \widetilde{\frak{m}}=
\frak l\circ\frak P \circ \widetilde{\frak{l}} + 2\frak C
\end{equation}
assuming that $\frak l$ is a unit-quaternion. 

Under consideration of \cite[Thm.\ 1]{ark2018} this representation implies that 
the one-parametric displacement can be represented by a straight line in the ambient space of the 
Study quadric. From \cite[Thm.\ 3]{ark2018} we can conclude immediately the following theorem:
\begin{thm}
The line-symmetric motion in $E^4$ with basic surface $\Gamma$  is a circular Darboux 2-motion\footnote{All 
points have circular trajectories.}, which is neither spherical nor a pure translation,  and vice versa.
\end{thm}

\section{Application}\label{sec:app}

As the Pl\"ucker  quadric $\Psi$ is a point-model of lines, one can use well-known methods for curves 
(freeform techniques, interpolation, approximation,\ldots) for the design of ruled surfaces. 
The challenge for applying this standard technique is that one has to deal with the side condition 
that the curve in $P^5$ has to be located on $\Psi$. 
For the task of interpolating given lines by a ruled surface one can apply for example the 
\begin{enumerate}[$\bullet$]
\item
rational interpolation of points on hyperquadrics \cite{gfrerrer_habil}, 
\item
interpolation with rational quadratic spline curves (biarc construction) \cite{wang_joe}, 
\cite[Sec.\ 2.12]{klawitter}. 
\end{enumerate}
Another possibility is to project the Pl\"ucker quadric stereographicly onto an affine 4-space $\AA^4$ 
\cite[p.\ 212]{pottmann_wallner}. Based on this approach a $G^1$-Hermite interpolation of 
ruled surfaces with low degree rational ruled surfaces is given in \cite{peternell}. 

Within this paper we do not focus on the interpolation problem but want to discuss the 
modification of the well-known algorithm of De Casteljau for the design of rational ruled surfaces. 
Applications in this context are e.g.\ 
wire cut EDM (electric discharge machining),
laser beam machining, cylindrical milling, 
generation of line-symmetric motions,  $\ldots$  

For the understanding of the following review on this topic we have to repeat a 
point-model for the set of oriented lines of $E^3$.

\subsection{Study sphere}\label{spears}

The two vectors $\Vkt l$ and $\mVkt l$ of Sec.\ \ref{sec:3space} can be combined by the so-called dual unit $\varepsilon$ with 
the property $\varepsilon^2=0$ to a dual vector $\dVkt l:=\Vkt l+\varepsilon \mVkt l$ with 
$\langle \dVkt l,\dVkt l\rangle \in\RR\setminus\left\{0\right\}$; i.e.\
\begin{equation}
\begin{split}
&\langle \dVkt l,\dVkt l\rangle = \langle \Vkt l+\varepsilon \mVkt l, \Vkt l+\varepsilon \mVkt l\rangle = \\
&\langle \Vkt l,\Vkt l\rangle + 2\varepsilon 
\langle \Vkt l,\mVkt l\rangle
=\langle \Vkt l,\Vkt l\rangle.
\end{split}
\end{equation}
Note that $\dVkt l$ is an element of $\DD^3$, where $\DD$ denotes the ring of dual numbers $a+\varepsilon b$ with 
$a,b\in\RR$. 

As $\Vkt l\neq \Vkt o$ holds one can additionally assume that $\Vkt l$ is a unit-vector; i.e.\ 
$\langle \Vkt l,\Vkt l\rangle=1$. As a consequence  $\dVkt l:=\Vkt l+\varepsilon \mVkt l$ is 
a so-called dual unit-vector representing a spear (oriented line). Therefore there is a bijection between the points of the 
dual unit-sphere $S_{\DD}^2\in\DD^3$ with
\begin{equation}
S_{\DD}^2:=\left\{
\dVkt l\in\DD^3 \quad \text{with} \quad \langle \dVkt l,\dVkt l\rangle=1
\right\}
\end{equation}
and the set of spears of $E^3$. Note that antipodal points of this so-called Study sphere 
correspond to oppositely oriented lines.

\subsection{Review}

On can think of the following possibilities for adapting 
De Casteljau's algorithm for the design of ruled surfaces using:
\subsubsection{A) Oriented Lines}
	\begin{enumerate}[$\star$]
	\item
	According to Odehnal \cite[Sec.\ 2]{odehnal_ruled} the quartic 
	manifold $M^4\in\RR^6$ given by 
	\begin{equation}
	M^4:\quad \langle\Vkt l,\Vkt l \rangle =1,\quad \langle\Vkt l,\mVkt l \rangle =0
	\end{equation}
	can be used	as point-model for the set of oriented lines. Then one can perform the  
	algorithm of De Casteljau in $\RR^6$ and project the resulting curve $(\Vkt c,\mVkt c)\in \RR^6$ back onto 
	$M^4$. This back projection $\theta$ is the composition of the mapping 
	$(\Vkt c,\mVkt c) \mapsto (\Vkt a,\mVkt a)$ according to Eq.\ (\ref{axis}) and the 
	normalization
	\begin{equation*}
	(\Vkt a,\mVkt a) \mapsto \left( \tfrac{\Vkt a}{\|\Vkt a\|}, \tfrac{\mVkt a}{\|\Vkt a\|} \right).
	\end{equation*} 
	This approach can be modified by projecting the points of $\RR^6$ obtained 
	after each iteration step of De Casteljau's algorithm back on $M^4$.
	\begin{rem}
	Analogue algorithms for oriented line-elements are given in \cite[Sec.\ 5.2.1]{nawratil_survey}. 
	\hfill $\diamond$
	\end{rem}
	\item
	The above described algorithm can also be performed within another setting; namely one can use as point-model 
	the dual unit-sphere $S_{\DD}^2$. Then De Casteljau's algorithm can be executed in $\DD^3$ and the 
	resulting curve is projected back onto $S_{\DD}^2$ by the normalization to dual unit-vectors. 
	This normalization is equivalent to the mapping $\theta$. This was in fact done by 
	Li and Ge \cite{li} for the generation of rational B\'{e}zier line-symmetric motions.
	\item
	Instead of these two projection algorithms one can think of a so-called geodesic one. It is based on the idea to 
	replace the straight lines of the control polygon in the ambient space of the used point-model by 
	their analog on the point-model, which are geodesics. Sprott and Ravani \cite{sprott} used as 
	point-model the dual unit-sphere $S_{\DD}^2$ where the geodesics (dual great circles) on $S_{\DD}^2$ 
	correspond to right helicoids\footnote{They are the only ruled minimal surfaces in $E^3$ 
	(and $E^4$; cf.\ \cite[p.\ 17]{plass}). A right helicoid is generated during a helical motion by a 
	line intersecting the axis orthogonally.}. Their resulting geodesic algorithm of De Casteljau on $S_{\DD}^2$ 
	was further studied in the context of light-weight concrete elements  \cite{hagemann}.
	
	In contrast to the projection algorithms the geodesic one has the disadvantage that the resulting 
	ruled surfaces are not rational, which is an important feature for the 
	interactive design of ruled surfaces in CAGD.
	\begin{rem}
	Geodesic algorithms for oriented line-elements are given in \cite[Sec.\ 5.2.2]{nawratil_survey}. 
	\hfill $\diamond$
	\end{rem}
	\end{enumerate}

\subsubsection{B) Un-oriented Lines}
	\begin{enumerate}[$\star$]
	\item
	One can use the already mentioned stereographic projection of the Pl\"ucker quadric $\Psi$ onto an affine 4-space $\AA^4$ 
	and perform De Casteljau's algorithm in $\AA^4$. Finally the obtained curve is stereographically projected back 
	to $\Psi$. The obtained ruled surface is rational but it depends on the chosen stereographic projection, which is less 
	satisfactory from the geometric point of view. 
	\item
	A very sophisticated method was presented by Ge and Ravani \cite{ge_ravani}. They identified antipodal points of 
	$S_{\DD}^2$ by considering the set of lines spanned by antipodal points, which is equivalent 
	to the projective plane over $\DD$. Within this projective dual plane they applied 
	a {\it projective De Casteljau algorithm}. 
	
	They studied the resulting rational ruled surfaces in detail and also determined surface patches 
	by using a second De Casteljau algorithm, which provides the distances of the boundary points of the patch along the 
	ruling from the striction point. Clearly this approach can easily be modified for the generation of 
	rational ruled surface strips. 
	\end{enumerate} 

In a projective De Casteljau algorithm (cf.\ \cite[Sec.\ 3]{pottmann_farin})  the control points come 
with weights, which are dual numbers in the case of \cite{ge_ravani}. 
But dual weights cannot be handled very intuitively by designers\footnote{Moreover the weights are not invariant under 
projective transformations.}. Ge and Ravani also mentioned that one can get 
rid of them by using so-called Farin points (which is also known as frame points or weight points), 
but they have not outlined a user-friendly method for interactive design. 
Based on our alternative interpretation such a method is presented in the next subsection. 

Moreover the straight forward extension to line-elements allows an analogue handling of 
ruled surface strips, which therefore can be generated by one projective De Casteljau algorithm instead of the 
combination of two algorithms as proposed in \cite{ge_ravani}. 

\begin{rem}\label{patch}
Note that our approach can easily be modified for B\'{e}zier ruled surface patches by 
replacing Eq.\ (\ref{hierl}) by
\begin{equation}\label{hierl_neu}
(l_{01}:l_{02}:l_{03}:l_{23}:l_{31}:l_{12}:l_1:l_2)
\end{equation}
with $l_i:=\langle \Vkt p_i,\Vkt l\rangle$ for $i=1,2$, where $\Vkt p_i$ are the two 
boundary points of the patch along the ruling. The non-homogenous version of this 
representation goes back to Ravani and Wang \cite{wang} and was furthered in 
\cite{ge_rav}.
\hfill $\diamond$
\end{rem}

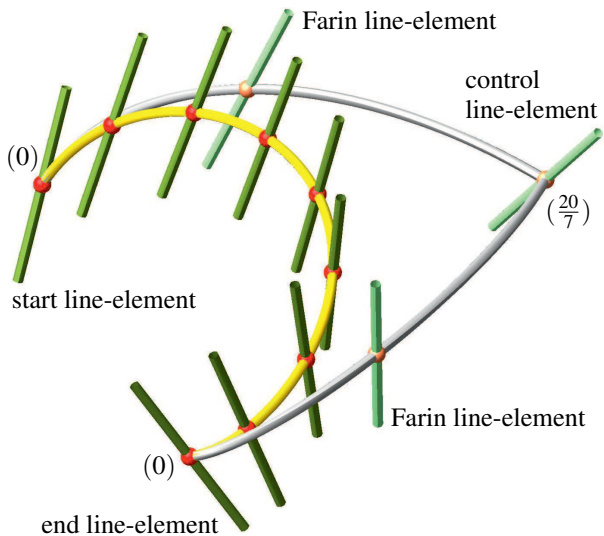
\begin{figure}[t]
\begin{center} 
 \begin{overpic}
    [height=70mm]{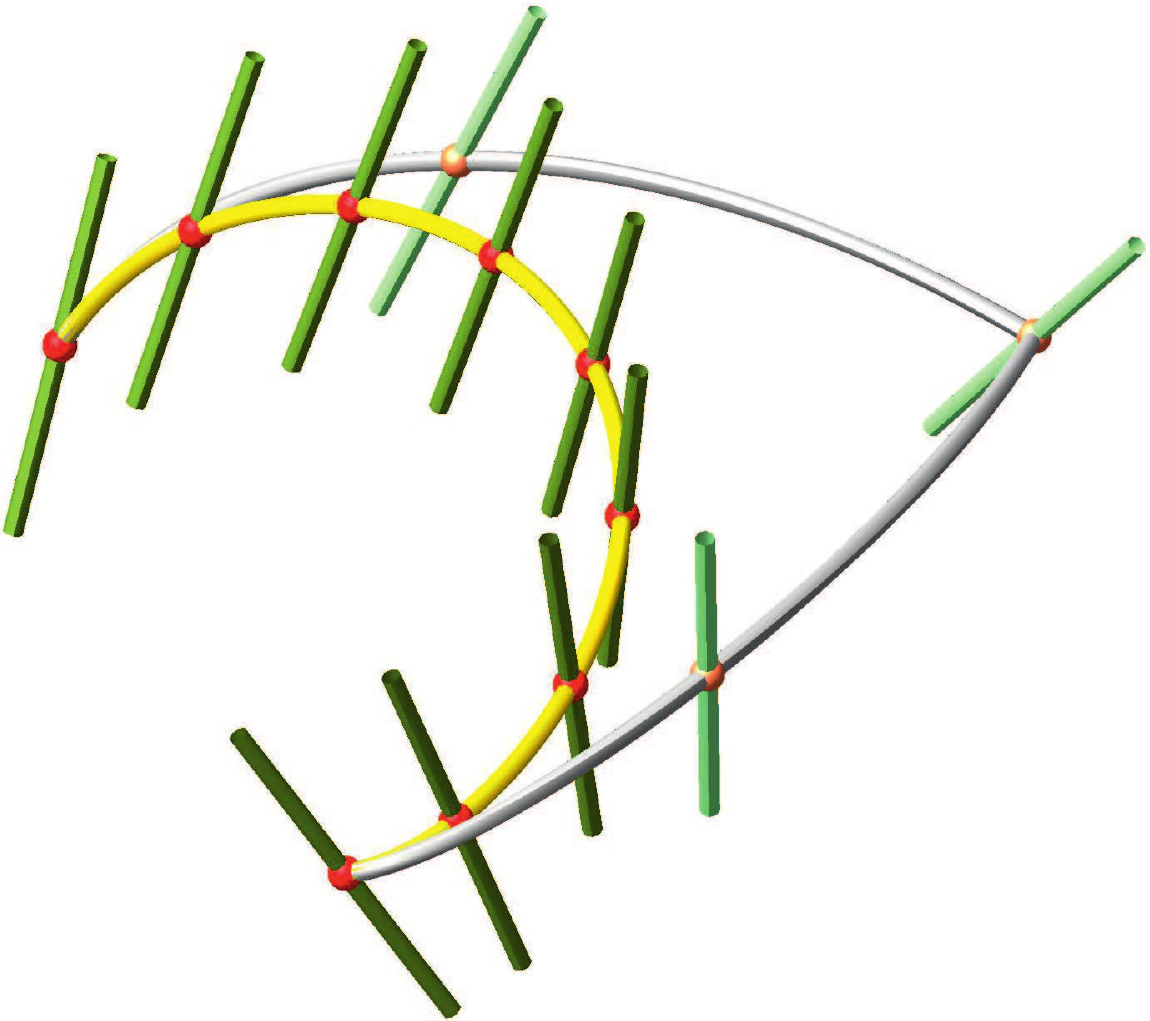}
\begin{footnotesize}
\put(22,10){$(0)$}
\put(5,0){end line-element}
\put(-1,62){$(0)$}
\put(0,38){start line-element}
\put(90,53){$(\tfrac{20}{7})$}
\put(76.5,75){control}
\put(76.5,70){line-element}
\put(49,85){Farin line-element}
\put(64,18){Farin line-element}
\end{footnotesize}      
  \end{overpic} 
\end{center} 
\caption{Example of the concluding algorithm: The design of a rational ruled surface strip 
is based on a very intuitively and user-friendly  control structure. 
 } 
\label{fig5}
\end{figure}

\subsection{Concluding algorithm}

In principle, the following algorithm is identical with the one of Ge and Ravani \cite{ge_ravani} 
but instead of working in the projective dual plane we use $P^5$ (for ruled surface strips/patches we 
work in $P^6$ and $P^7$, respectively). In this way we  gain a better geometric understanding, 
which results in a user-friendly control. The procedure is as follows:

We perform a projective De Casteljau algorithm in the projective space of dimension 5 (6 and 7, respectively),  
which uses Farin points instead of weights. The resulting curve can be interpreted as a 
conoidal ruled 2-surface (strip/patch) in $E^4$ with respect to the director hyperplane $x_0=0$ (cf.\ Sec.\ \ref{sec:line4} and \ref{sec:lineele4}). 
By applying the orthogonal projection $\pi$ in $x_0$-direction we obtain the desired ruled surface (strip/patch) in $E^3$. 
Moreover we label the projected lines (line-elements/line-segments) by the $x_0$-coordinate 
("{\it kotierte Projektion}", cf.\ Fig.\ \ref{fig2}). In this way the user can modify very 
intuitively the control structure; i.e.\ the Farin and control lines (line-elements/line-segments) can be changed by {\it mouse action}
and their $x_0$-heights by the {\it scroll wheel}. 

In Fig.\ \ref{fig5} we illustrated a quartic rational ruled surface strip, which  corresponds to a quadratic Bezier curve in $P^6$. 
Each {\it Farin line-element} can only be modified within the ruled surface strip (composed of a Pl\"ucker conoid and an 
ellipse on it; cf.\ Sec.\ \ref{sec:straight}) determined by the {\it control line-element} and {\it start/end line-element}, respectively. 
In contrast the {\it control line-element} has $6$ degrees of freedom. 
The $x_0$-values of the control, start and end line-element are given in parentheses.

Finally it should be noted that an analogue algorithm for the design of rational motions in $E^3$ is given in \cite{ark2018}.
 
\begin{rem}
The rational surface obtained by our algorithm can be written in B\'{e}zier representation.
A geometric interpretation of the corresponding Farin lines and control lines in terms of 
linear complexes is given in \cite{ppr}. \hfill $\diamond$
\end{rem}

\section*{Acknowledgments}
 
The author is supported by Grant No.~P~24927-N25 of the Austrian Science Fund FWF.

%%% You can use BibTeX for formatting your references:
%\bibliography{icgg}

\begin{thebibliography}{99}
%
\bibitem{bottemaroth}
O.~Bottema and B.~Roth: \newblock
\emph{Theoretical Kinematics}. \newblock North-Holland Publishing Company, Amsterdam, 1979.
%
\bibitem{ge_rav}
Q.J.~Ge and B.~Ravani: \newblock 
On representation and interpolation of line-segments for computer aided geometric design. \newblock
\emph{ASME Adv.\ Design Autom.}, 69(1):191--198, 1994.
%
\bibitem{ge_ravani}
Q.J.~Ge and B.~Ravani: \newblock 
Geometric Design of Rational B\'{e}zier Line Congruences and Ruled Surfaces Using Line Geometry. \newblock 
\emph{Computing Supplement 13}, pages 101--120. Springer-Verlag, New York, 1998. 
Edited by G.E.~Farin.
%
\bibitem{gfrerrer_habil}
A.~Gfrerrer: \newblock
On the construction of rational curves on hyperquadrics. \newblock
Habilitation thesis, Graz University of Technology, 2001.
%
\bibitem{hagemann}
M.~Hagemann and D.~Klawitter: \newblock 
Discretisation of light-weight concrete elements using a line-geometric model. \newblock 
In \emph{Proceedings of the 9th fib International PhD Symposium in Civil Engineering 
(Karlsruhe Institute of Technology, Germany, July 22--25)}, pages 269--274. 
KIT Scientific Publishing, Karlsruhe, 2012. Edited by H.S.~M\"uller et al.
 %
\bibitem{hopsw}
M.~Hofer, B.~Odehnal, H.~Pottmann, T.~Steiner and J.~Wallner: \newblock 
3D shape recognition and reconstruction based on line element geometry.  \newblock 
In \emph{Proceedings of the Tenth IEEE International Conference on Computer Vision - Volume 2 
(Beijing, China, October 17--20)}, pages 1532--1538. IEEE Computer Society Washington, DC, 2005.
%
\bibitem{klawitter}
D.~Klawitter: \newblock 
\emph{Clifford Algebras -- Geometric Modelling and Chain Geometries with Application in 
Kinematics}. \newblock Springer Spektrum, Wiesbaden, 2015.
%
\bibitem{li}
S.~Li and Q.J.~Ge: \newblock 
Rational B\'{e}zier Line-Symmetric Motions. \newblock 
\emph{J.\ Mech.\ Des.}, 127(2):222--226, 2005.
% 
\bibitem{nawratil_clifford}
G.~Nawratil: \newblock
Fundamentals of quaternionic kinematics in Euclidean 4-space. \newblock
\emph{Adv.\ Appl.\ Clifford Algebras}, 26(2):693--717, 2016. 
%
\bibitem{nawratil_line_element}
G.~Nawratil: \newblock
Quaternionic approach to equiform kinematics and line-elements of Euclidean 4-space and 3-space. \newblock
\emph{Comput.\ Aided Geom.\ Des.}, 47:150--162, 2016. 
%
\bibitem{nawratil_survey}
G.~Nawratil: \newblock
Point-models for the set of oriented line-elements – a survey. \newblock
\emph{Mech.\ Mach.\ Theory}, 111:118--134, 2017.
%
\bibitem{ark2018}
G.~Nawratil: \newblock 
Kinematic interpretation of the Study quadric's ambient space. \newblock 
In \emph{Proceedings of the 16th International Symposium on Advances in Robot Kinematics  
(Bologna, Italy, July 1--5)}, accepted. Springer, 2018. 
arXiv:1708.02622, 2017.
%
\bibitem{opw}
B.~Odehnal, H.~Pottmann and J.~Wallner: \newblock 
Equiform kinematics and the geometry of line elements. \newblock 
\emph{Beitr.\ Algebra Geom.}, 47(2):567--582, 2006.
%
\bibitem{boris}
B.~Odehnal: \newblock 
Die Linienelemente des $P^3$. \newblock 
\emph{\"Osterreich.\ Akad.\ Wiss.\ Math.-Naturw.\ Kl.\ S.-B.\ II}, 215:155--171, 2006. 
%
\bibitem{odehnal_ruled}
B.~Odehnal: \newblock 
Subdivision Algorithms for Ruled Surfaces. \newblock 
\emph{J. Geom.\ Graphics}, 12(1):1--18, 2008.
%
\bibitem{ppr}
M.~Peternell, H.~Pottmann and B.~Ravani: \newblock 
On the computational geometry of ruled surfaces. \newblock 
\emph{Comput.\ Aided Des.}, 31:17--32, 1998.
%
\bibitem{peternell}
M.~Peternell: \newblock 
$G^1$-Hermite Interpolation of Ruled Surfaces. \newblock 
\emph{Mathematical Methods in CAGD: Oslo 2000}, pages  413--422.  Vanderbilt Univ.~Press, Nashville, TN, 2001. 
Edited by T.~Lyche and L.L.~Schumaker.
%
\bibitem{martin}
M.~Peternell and B. ~Odehnal: \newblock 
On Generalized LN-Surfaces in 4-Space. \newblock
In \emph{Proceedings of the International Symposium on Symbolic and Algebraic Computation
(Linz/Hagenberg, Austria, July 20--23)}, pages 223--230. 
ACM, 2008. Edited by J.R.~Sendra and L.~Gonzalez-Vega.
%
\bibitem{without_study}
M.~Pfurner, H.-P.~Schr\"ocker and M.~Husty: \newblock
Path Planning in Kinematic Image Space Without the Study Condition. \newblock 
\emph{Advances in Robot Kinematics 2016}, pages 285--292. Springer, 2018.
Edited by J.~Lenarcic and J.-P.~Merlet.
%
\bibitem{plass}
M.H.~Plass: \newblock
Ruled surfaces in Euclidean four space. \newblock
Doctoral thesis, Massachusetts Institute of Technology, 1939.
%
\bibitem{pottmann_farin}
H.~Pottmann and G.~Farin:  \newblock
Developable rational  B\'{e}zier and B-spline surfaces. \newblock
\emph{Comput.\ Aided Geom.\ Des.}, 12:513--531, 1995.
%
\bibitem{pottmann_wallner}
H.~Pottmann and J.~Wallner: \newblock 
\emph{Computational Line Geometry}. \newblock Springer, Berlin Heidelberg, 
2001.
%
\bibitem{wang}
B.~Ravani and J.W.~Wang: \newblock
Computer Aided Geometric Design of Line Constructs. 
\emph{J.\ Mech.\ Des.}, 113:363--371, 1991.
%
\bibitem{swc}
J.M.~Selig,Y.~Wu and M.~Carricato: \newblock
Motion Interpolation in Lie Subgroups and Symmetric Subspaces. \newblock
\emph{Computational Kinematics}, pages 467--474. Springer, 2017.
Edited by S.~Zeghloul et al. 
%
\bibitem{sprott}
K.~Sprott and B.~Ravani: \newblock 
Kinematic generation of ruled surfaces. \newblock 
\emph{Adv.\ Comp.\ Math.} 17:115--133, 2002. 
%
\bibitem{wang_joe}
W.~Wang and B.~Joe: \newblock 
Interpolation on quadric surfaces with rational quadratic spline curves. \newblock 
\emph{Comput.\ Aided Geom.\ Des.}, 14(3):207--230, 1997.
%
 \end{thebibliography}
%%% If you prefer to format your references manually, do it as
%%% follows:

% Every article should conclude with some basic information about the
% authors:
\section*{About the author}
Georg Nawratil is a senior researcher 
in the research group "Differential Geometry and Geometric Structures"  
at the Institute of Discrete
Mathematics and Geometry, Vienna University of Technology, Austria. 
Moreover he is a member of the IFToMM Technical Commitee for "Computational Kinematics"
and of the "Center for Geometry and Computational Design", Vienna University of Technology, Austria. 
His research interests 
include the geometry of mechanisms, kinematics, robotics and line geometry. 
He can be reached by email:
nawratil@geometrie.tuwien.ac.at, 
by phone: +43-1-58801-104362,
by fax: +43-1-58801-9104362, 
or through the postal address:
Institute of Discrete Mathematics and Geometry,
Vienna University of Technology, Wiedner
Hauptstrasse 8-–10/104, A-1040 Vienna, Austria, Europe.
More informations about the author can be found on his homepage:      
http://www.geometrie.tuwien.ac.at/nawratil/. \balance

%\noindent{\em Note: if possible, make the length of the two columns balanced in case your text does not fill the whole page, in order to have a better ‘last page’ 5%in the Full Paper. For instance, use the \verb|\balance| instruction anywhere in the text that originally appears on the last left column} \balance

\end{document}